\documentclass{iopart}

\usepackage{graphicx}

\usepackage[hang]{subfigure}

\DeclareSymbolFont{largesymbols}{OMX}{txex}{m}{n}

\begin{document}

\title{A tomographic approach to quantum nonlocality}

\author{Stefano Mancini$^\dag$, Vladimir I. Man'ko$^\ddag$, \\
E. V. Shchukin$^\ddag$ and Paolo Tombesi$^\dag$}

\address{\dag\ INFM, Dipartimento di Fisica, Universit\`a di Camerino, 
I-62032 Camerino, Italy}

\address{\ddag\ P.N. Lebedev Physical Institute, Leninskii Prospect 53, 
Moscow 119991, Russia}

\date{\today}

\begin{abstract}
We propose a tomographic approach to study quantum nonlocality 
in continuous variable quantum
systems. On one hand we derive a Bell-like inequality for measured 
tomograms. On the other
hand, we introduce pseudospin operators whose statistics can be 
inferred from the data
characterizing the reconstructed state,
thus giving the possibility to use standard Bell's
inequalities. Illuminating examples are also discussed.
\end{abstract}


{\bf keywords}: entanglement, nonlocality, quantum tomography, 
measurement theory.

\maketitle

\section{Introduction}

In their famous paper of 1935 \cite{EPR}, Einstein, Podolsky, and Rosen (EPR) 
questioned the
completeness of quantum mechanics (QM). Their argument was based on the 
premise of no action
at a distance (locality) and realism. Several years later Bell \cite{BEbook} 
showed that the
prediction of QM are incompatible with the premises of local realism 
(or local hidden variable
theories).

Experiments \cite{EXP} based on Bell's result support QM, 
indicating the failure of local
hidden variable theories. Quantitative tests used mainly 
Bohm's (dichotomic)  version
\cite{BO} of the EPR entangled states instead of the 
original EPR states with continuous
degrees of freedom.

In recent years, systems with continuous variable has 
attracted much atttention in connection
with the burgeoning field of quantum information \cite{QI}. 
In such a field EPR entanglement
and quantum nonlocality are of practical importance. 
Nevertheless, the generalization of
Bell's inequalities to quantum systems with continuous 
variabbles still remains a challenging
issue.

In continuous variable bipartite quantum systems, 
EPR aspects can arise when trying to infer
quadratures of one subsystem from those of the other 
\cite{RD}. These can be tested by
exploiting the Heisenberg uncertainty principle \cite{RD}. 
Such approach has been employed in Refs.\cite{OU,GMT}. 
Few other theoretical proposals use the field quadrature measurements 
\cite{GIL,MUN,WEN}, while a phase space approach to quantum nonlocality 
has been developed \cite{BW} and tested \cite{KUZ}.

On the other hand quantum tomography \cite{TOMOS} 
provides a useful tool to reach a complete
state knowledge. Such knowledge could 
then be used for nonlocal tests. Some aspects of EPR
problem have been considered in tomographic approach in \cite{AM}.

Here, we propose to use the tomographic data to 
do nonlocal tests. At first instance, we shall
develop an inequality based on the tomograms. 
However, this will result quite difficult to
violate. A state does not have to violate all possible 
Bell's inequalities to be considered
quantum nonlocal; rather, a given state is nonlocal 
when it violates any Bell's inequality
\cite{JEONG}. Thus the degree of quantum nonlocality 
that we can uncover crucially depends not
only on the given quantum state, but also on the 
Bell operator \cite{BMR}. Then, we shall
introduce pseudospin operators whose statistics 
can be inferred from the data characterizing
the reconstructed state, thus giving the possibility 
to use standard Bell's inequalities. We
shall also present some illuminating examples.

\section{Quantum tomography revisited}

A quantum state is described by the density operator 
$\rho$ or by any its phase space
representation like the Wigner 
function $W(q,p)$, where $q$, $p$ are canonical conjugate
variables. Quantum tomography is a technique which allows 
to recover the quantum state from a
set of measured probability distributions. 
The latter, sometimes called tomograms, are line
projections of the Wigner function. Such projections 
can be parametrized through a symplectic
transform \cite{SYMPL}. Practically, the 
tomograms can be expressed as
\begin{equation}\label{eq:wW}
    w(X,\mu,\nu)=\int W(q,p)\delta(X-q\mu-p\nu)\,dq\,dp\,,
\end{equation}
where $\mu,\;\nu\in {\bf R}$, are the parameters of the 
symplectic transformation and $X$ is a
stochastic variable. It represents the random outcomes of 
the measurement of the observable
$X=\mu q+\nu p$, and $w$ results the probability distribution 
associated to such observable.
It is worth noting that the above symplectic transformation, 
once thought as composition of
rotations and squeezing, can be realized in optical systems 
\cite{JMO}. A remarkable property
of the tomograms (\ref{eq:wW}) is their homogeneity.

To get the phase space picture of the quantum state one has to 
invert the relation
(\ref{eq:wW}), obtaining
\begin{equation}\label{eq:Ww}
    W(q,p)=\int w(X,\mu,\nu) e^{-i\mu q-i\nu p+iX}\,dX\,d\mu\,d\nu \,.
\end{equation}
This simply results an inverse Fourier transform. 
Neverthless, it becomes a more involved
inverse transform, namely an inverse Radon transform, 
when reducing the symplectic
transformation to a mere rotation, i.e., 
$\mu=\cos\theta$, $\nu=\sin\theta$. This is the case
of optical homodyne tomography (OHT).

By expanding the density operator in terms of a 
complete set of operators, it is also be
possible to directly relate the quantum state to the tomograms, that is
\begin{equation}\label{eq:rhow}
    \rho=\int w(X,\mu,\nu)K(X,\mu,\nu)\,dX\,d\mu\,d\nu\,,
\end{equation}
where the kernel operator takes the form
\begin{equation}\label{eq:calK}
    K(X,\mu,\nu)=\frac1{2\pi}\lambda^2
    \exp\left(-i\lambda X+i\lambda^2\frac{\mu\nu}2\right)
    \exp(i\lambda\mu q)\exp(i\lambda\nu p)\,.
\end{equation}
Here, $\lambda$ can be set equal to unity; this freedom 
reflects the overcompleteness of the
information obtainable by means of all possible tomograms (\ref{eq:wW}).

In OHT it is preferable to use Eq. (\ref{eq:rhow}) instead of 
Eq. (\ref{eq:Ww}) since the
kernel in the latter case is unbounded, 
thus requiring the use of an artificial cutoff to
directly sample the Wigner function.
On the contrary, Eq. (\ref{eq:rhow}) allows a direct
sample of the density matrix elements on the basis 
where $K$ results bounded \cite{DAR}.

\section{Bell-like inequality for tomograms}

The above arguments can be generalized to the 
case of a bipartite two-mode system
\cite{TWOMODE}. In such a case the tomograms will be given by
\begin{eqnarray}\label{eq:wW2}
    w(X_1,\mu_1,&\nu_1&,X_2,\mu_2,\nu_2) = \nonumber
    \int dq_1dp_1 \int W(q_1,p_1,q_2,p_2) \times \\
    &\times& \delta(X_1-q_1\mu_1-p_1\nu_1)
    \delta(X_2-q_2\mu_2-p_2\nu_2)\,dq_2dp_2\,,
\end{eqnarray}
where $\mu_1$, $\nu_1$, $\mu_2$, $\nu_2$ 
are parameters. Let us restrict to the case of OHT,
and suppose to perform homodyne detection 
in both subsystems $1$ and $2$. Then,
$\mu_1=\cos\theta_1$, $\nu_1=\sin\theta_1$, 
$\mu_2=\cos\theta_2$, $\nu_2=\sin\theta_2$, where
$\theta_1$, $\theta_2$ are the rotation angles 
related to the local oscillators phase. As a
consequence we have 
$w(X_1,\mu_1,\nu_1,X_2,\mu_2,\nu_2)$ $\to w(X_1,\theta_1,X_2,\theta_2)$.

We now classify the results 
of the measurements to be $+$ if the quadrature result $X$ is
greater than zero, and $-$ otherwise \cite{GIL}. 
Then, we can construct the following
probabilities from the tomograms
\begin{eqnarray}\label{eqs:wpm}
    w_{++}(\theta_1,\theta_2) &=& 
    \int\limits_0^{\infty}\,dX_1\,\int\limits_0^{\infty}
    w(X_1,\theta_1,X_2,\theta_2)\,dX_2\,, \\
    w_{+-}(\theta_1,\theta_2) &=& 
    \int\limits_0^{\infty}\,dX_1\,\int\limits_{-\infty}^0
    w(X_1,\theta_1,X_2,\theta_2)\,dX_2\,, \\
    w_{-+}(\theta_1,\theta_2) &=& 
    \int\limits_{-\infty}^0\,dX_1\,\int\limits_0^{\infty}
    w(X_1,\theta_1,X_2,\theta_2)\,dX_2\,, \\
    w_{--}(\theta_1,\theta_2) &=& 
    \int\limits_{-\infty}^0\,dX_1\,\int\limits_{-\infty}^0
    w(X_1,\theta_1,X_2,\theta_2)\,dX_2\,.
\end{eqnarray}
Then, the Bell's inequalities can be written 
in terms of the above probabilities. In
particular, the CHSH inequality \cite{CHSH} can be written as
\begin{equation}\label{eq:Bdef}
    B\equiv\left|E(\theta_1,\theta_2)+
    E(\theta_1,\theta_2')+
    E(\theta_1',\theta_2)-
    E(\theta_1',\theta_2')\right|\le 2\,,
\end{equation}
where now
\begin{equation}\label{eq:Edef}
    E(\theta_1,\theta_2)=
    w_{++}(\theta_1,\theta_2)-
    w_{+-}(\theta_1,\theta_2)-
    w_{-+}(\theta_1,\theta_2)+
    w_{--}(\theta_1,\theta_2).
\end{equation}

\section{Reconstructed data and pseudo-Spin operators}

As discussed in Section 2, measuring tomograms allows 
to statistically sample the density
matrix elements at least in some basis, e.g. Fock basis. 
This represents a complete knowledge
of the state, also for a bipartite system. 
Then, such information, say the density matrix
elements in Fock basis, can be used to derive 
the statistics of any hermitian operator, even
if this is not directly measurable. The price 
one has to pay is the large amount of data needs
to collect.

Let us now see how these arguments can be fruitful 
applied to the problem of quantum
nonlocality. We introduce the local pseudo-spin 
operators \cite{FIU}
\begin{eqnarray}\label{eq:S}
    S_{x}^{(j)} &=& \sum_{n=0}^{\infty}\Bigl(|2n \rangle_{j}\langle 2n+1|
    +|2n+1 \rangle_{j}\langle 2n|\Bigr)\,, \nonumber \\
    S_{y}^{(j)} &=& -i\sum_{n=0}^{\infty}\Bigl(|2n \rangle_{j}\langle 2n+1|
    -|2n+1 \rangle_{j}\langle 2n|\Bigr)\,, \\
    S_{z}^{(j)} &=& \sum_{n=0}^{\infty}(-)^{n}
    |n \rangle_{j}\langle n|\,, \nonumber
\end{eqnarray}
where $j=1,2$ and $|n\rangle_{j}$ are Fock states. 
The operators (\ref{eq:S}) obey the
commutation relation of spin-$\frac{1}{2}$ algebra, namely
\begin{equation}\label{eq:SCR}
    \left[S_{x}^{(j)},S_{y}^{(j')}\right]
    =2\,i\,\delta_{j,j'}\;\varepsilon_{xyz}\,S_{z}^{(j)}\,,
\end{equation}
with $\varepsilon_{xyz}$ the totally antisymmetric tensor. 
Spin tomography approach was
considered e.g. in \cite{DodPLA}.

In terms of the operators (\ref{eq:S}) the CHSH 
inequality \cite{CHSH} reads in its standard
form, that is
\begin{equation}\label{eq:calB}
    \mathcal{B}\equiv|\mathcal{E}({\bf u},{\bf v})
    +\mathcal{E}({\bf u}',{\bf v})
    +\mathcal{E}({\bf u},{\bf v}')-\mathcal{E}({\bf u}',{\bf v}')|\le 2\,,
\end{equation}
with
\begin{equation}\label{eq:calE}
    \mathcal{E}({\bf u},{\bf v})=
    \Big\langle\left({\bf u}\cdot{\bf S}^{(1)}\right)
    \left({\bf v}\cdot{\bf S}^{(2)}\right)\Big\rangle\,.
\end{equation}
Here ${\bf u}$, ${\bf u}'$, ${\bf v}$, ${\bf v}'$ 
are unit vectors in ${\bf R}^3$, while ${\bf S}^{(j)}
\equiv(S_{x}^{(j)},S_{y}^{(j)},S_{z}^{(j)})$ 
and the dot indicates the ordinary scalar
product in ${\bf R}^3$. Furthermore, the angle 
brackets in Eq. (\ref{eq:calE}) denote the
average over the density matrix elements 
assumed available from tomographic reconstruction.

Let us now study the possible violations of inequality 
(\ref{eq:calB}) for the two preceeding
examples. For simplicity, we consider the directions 
${\bf u}$, ${\bf u}'$, ${\bf v}$, ${\bf v}'$ coplanar, 
in the plane $x-z$, so that
\begin{equation}\label{eq:calExz}
    \mathcal{E}({\bf u},{\bf v})=\Big\langle
    \left(S_{z}^{(1)}\cos\theta_{u}
    +S_{x}^{(1)}\sin\theta_{u}\right)
    \left(S_{z}^{(2)}\cos\theta_{v}+
    S_{x}^{(2)}\sin\theta_{v}\right)
    \Big\rangle\,,
\end{equation}
where $\theta_{u}$ ($\theta_{v}$) is the 
relative angle betweeen ${\bf u}$ (${\bf v}$) and
${\bf z}$.

\section{Examples}

Let us going to study the possible violations 
of the inequalities (\ref{eq:Bdef}) and
(\ref{eq:calE}) in three paradigmatic cases.

\subsection{Example A}

We first consider the two-mode squeezed vacuum state 
which can be generated in the
non-degenerate optical parametric amplifier \cite{QObook}
\begin{equation}\label{eq:PsiEPR}
    |\psi\rangle=\sqrt{1-\lambda^2}\sum_{n=0}^{\infty}
    \lambda^n |n\rangle_1|n\rangle_2\,,
\end{equation}
where $\lambda=\tanh s$ and $s$ is the squeezing parameter. 
Such state can also be described
by the following Wigner function
\begin{eqnarray}\label{eq:WEPR}
    W(q_1,p_1,q_2,p_2) &=& \frac4{\pi^2}\exp\Bigl(
    -e^{-2s}\left[(q_1-q_2)^2+(p_1+p_2)^2\right] \Bigr. \nonumber \\
    &-&\Bigl.e^{2s} \left[(q_1+q_2)^2+(p_1-p_2)^2\right]\Bigr)\,.
\end{eqnarray}
Notice that for $s\to\infty$ the state (\ref{eq:WEPR}) 
becomes precisely the EPR state.

Inserting Eq. (\ref{eq:WEPR}) into Eq. (\ref{eq:wW2}) we obtain
\begin{equation}\label{eq:wEPR}
    w(X_1,\theta_1,X_2,\theta_2)=\frac2{\pi}N
    \exp\left(-2aX_1^2-2aX_2^2-4aX_1X_2\right)\,,
\end{equation}
where
\begin{eqnarray}\label{eqs:abN}
    a &=& \frac{\cosh(2s)}{\cosh^2(2s)
    -\sinh^2(2s)\cos^2(\theta_1+\theta_2)}\,, \nonumber \\
    b &=& \frac{\sinh(2s)\cos(\theta_1+\theta_2)}{\cosh^2(2s)-\sinh^2(2s)
         \cos^2(\theta_1+\theta_2)}\,, \\
    N &=& \sqrt{a^2-b^2}\,. \nonumber
\end{eqnarray}
In this case, the tomograms depend only on the sum of the parameters.

Using the position of Eqs. (\ref{eqs:wpm}) we obtain
\begin{eqnarray}
    w_{++}(\theta_1,\theta_2) &=& w_{--}(\theta_1,\theta_2)
    =\frac1{4\pi}\left[\pi-2\arctan\left(\frac{b}{N}\right)\right]\,, 
    \label{eqs:wppEPR} \\
    w_{+-}(\theta_1,\theta_2) &=& w_{-+}(\theta_1,\theta_2)
    =\frac1{4\pi}\left[\pi+2\arctan\left(\frac{b}{N}\right)\right]\,, 
    \label{eqs:wpmEPR}
\end{eqnarray}
where the signs in front of $\pi$ refers to 
$\pi/2\le(\theta_1+\theta_2)\le\pi/2$, and should
be adjusted accordingly to the range of $(\theta_1+\theta_2)$.

\begin{figure}
\begin{center}
    \subfigure[The quantity $w_{++}(\theta_1,\theta_2)$ 
    of Eq. (\ref{eqs:wpmEPR}) is shown vs
               $(\theta_1+\theta_2)$ for 
	       different values of $\lambda$. The quantity
               $w_{+-}(\theta_1,\theta_2)$ has identical 
	       behavior with an offset of $\pi$ in
               $(\theta_1+\theta_2)$.]{\includegraphics{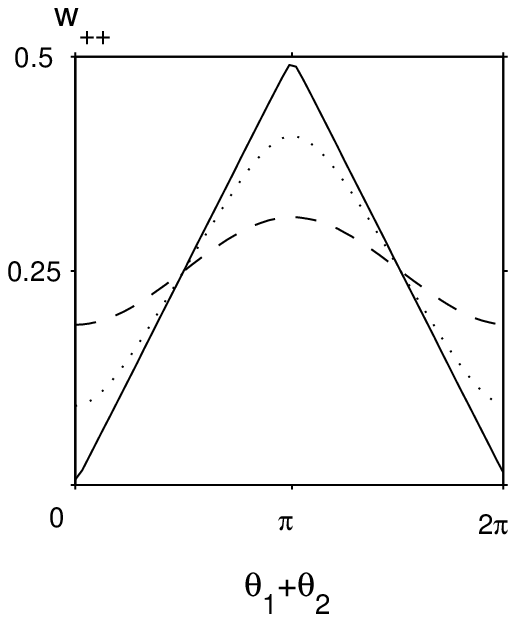} 
	       \label{fig1a}}
    \hspace{5mm}
    \subfigure[The quantity $\mathcal{B}$ of 
    Eq. (\ref{eq:calEEPR}) is shown vs $\theta_{u}$ for
               different values of $\lambda$. The other 
	       parameters are $\theta_{v}=\pi/4$,
               $\theta_{u'}=-\pi/2$ and $\theta_{v'}=-\pi/4$.]
	       {\includegraphics{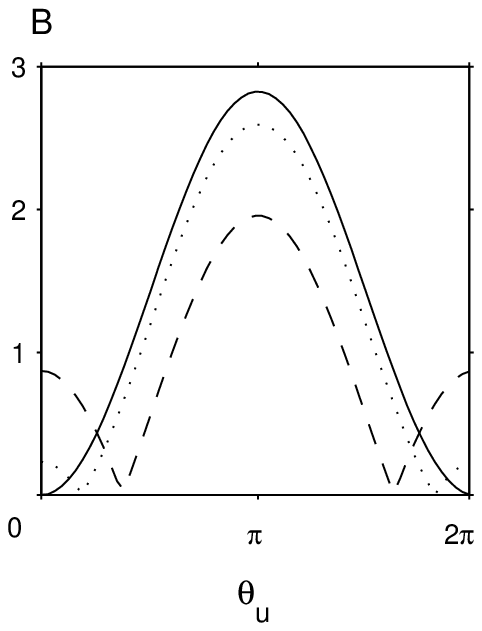}
               \label{fig1b}}
\end{center}
\caption{Example A (dashed line $\lambda=0.20$, 
dotted line $\lambda=0.54$, solid line
$\lambda=0.96$)}
\end{figure}

In Fig.\ \ref{fig1a} we show the behavior of the functions
(\ref{eqs:wppEPR})-(\ref{eqs:wpmEPR}) vs $\theta_1+\theta_2$ 
for different values of
$\lambda$. As we can see they never exceed 
the value $1/2$, thus the quantities $E$, in Eq.
(\ref{eq:Edef}), will be always confined between 
$1$ and $-1$. In turns, the inequality
(\ref{eq:Bdef}) will be never violated. 
This results seems in agreement with Bell's
conclusions stating that the original EPR 
state does not exhibit nonlocality because its
Wigner function is positive everywhere, and as 
such allows a hidden variable description
\cite{BEbook}. However, we shall see that 
is simply a matter of the type of measurements.

For the state (\ref{eq:PsiEPR}), Eq. (\ref{eq:calExz}) becomes
\begin{equation}\label{eq:calEEPR}
    \mathcal{E}({\bf u},{\bf v}) = \cos\theta_{u}\cos\theta_{v}+
    \frac{2\lambda}{1+\lambda^2}\sin\theta_{u}\sin\theta_{v}\,.
\end{equation}
Then, in Fig. \ref{fig1b} we show $\mathcal{B}$ for different 
values of $\lambda$. In this
case the EPR state show its nonlocality. 
Indeed a maximal violation of inequality
(\ref{eq:calB}) is achieved for $\lambda\to 1$ ($s\to\infty$).

\subsection{Example B}

As second example we consider the entangled superposition
\begin{equation}\label{eq:PsiN}
    |\psi\rangle=\frac1{\sqrt2}\left(|0\rangle_{1}|0\rangle_{2}
    +|n\rangle_{1}|n\rangle_{2}\right)\,.
\end{equation}
Since $|0\rangle$ and $|n\rangle$ are orthogonal for $n\ne 0$, 
the state (\ref{eq:PsiN})
resembles a spin triplet state. The corresponding Wigner function is
\begin{eqnarray}\label{eq:WN}
    W(q_1,p_1,q_2,p_2) &=& \frac{1}{2\pi^2}
    \biggl(1+\frac{2^n}{n!}{(q_1-ip_1)}^n{(q_2-ip_2)}^n+ \biggr. 
    \nonumber \\
    &+&\frac{2^n}{n!}{(q_1+ip_1)}^n{(q_2+ip_2)}^n+ \\
    &+&\biggl.L_n(2q^2_1+2p^2_1)L_n(2q^2_2+2p^2_2)\biggr)
    \,e^{-q^2_1-q^2_2-p^2_1-p^2_2}.
    \nonumber
\end{eqnarray}
where $L_{n}$ indicates the Laguerre polynomial of order $n$. 
By means of Eqs. (\ref{eq:wW2})
and (\ref{eq:WN}) we get the tomogram of the state under consideration
\begin{eqnarray}\label{eq:wN}
    w(X_1,\theta_1,X_2,\theta_2) &=& \frac1{2\pi}
    \biggl(1+\frac{H_n(X_1)H_n(X_2)}{2^{n-1}n!}
    \cos n(\theta_1-\theta_2)+\biggr. \nonumber \\
    &+&\biggl.\frac{H^2_n(X_1)H^2_n(X_2)}{2^{2n}{(n!)}^2}\biggr)
    e^{-X^2_1-X^2_2}.
\end{eqnarray}
where $H_n$ indicates the Hermite polynomial of order $n$. 
Using the position of Eqs.
(\ref{eqs:wpm}) we obtain probabilities $w_{\pm\pm}(\theta_1, \theta_2)$
\begin{eqnarray}\label{eq:wpmN}
    w_{++}(\theta_1,\theta_2) &=& w_{--}(\theta_1,\theta_2) =
    \frac{1}{4}+\frac{H^2_{n-1}(0)}{\pi 2^nn!}\cos n(\theta_1+\theta_2), \\
    w_{+-}(\theta_1,\theta_2) &=& w_{-+}(\theta_1,\theta_2) =
    \frac{1}{4}-\frac{H^2_{n-1}(0)}{\pi 2^nn!}\cos n(\theta_1+\theta_2).
\end{eqnarray}

\begin{figure}
\begin{center}
    \subfigure[The quantity $w_{++}(\theta_1,\theta_2)$ 
    of Eq. (\ref{eq:wpmN}) is shown vs
               $(\theta_1+\theta_2)$ for different values of $n$ 
	       (dashed line $n=1$,
               dotted line $n=3$, solid line $n=5$).]
               {\includegraphics{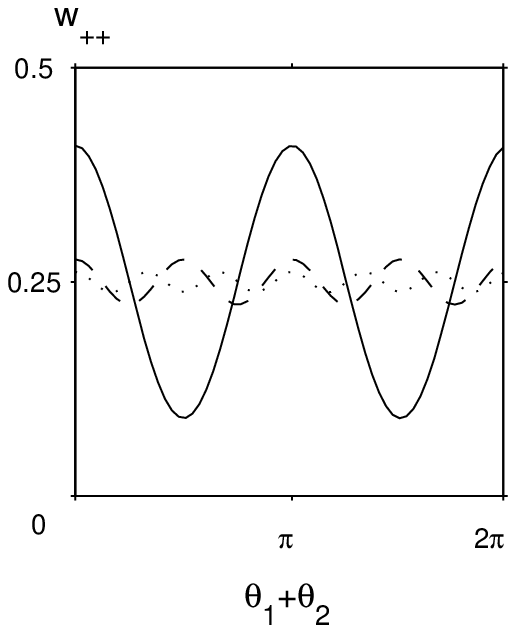} \label{fig2a}}
    \hspace{5mm}
    \subfigure[The quantity ${\cal B}$ derived from Eq. (\ref{eq:EB})
               is shown vs $\theta_{u}$ for case of $n=1$. 
	       The other parameters are
               $\theta_{v}=0$, $\theta_{u'}=\pi$ and $\theta_{v'}=\pi/2$.]
               {\includegraphics{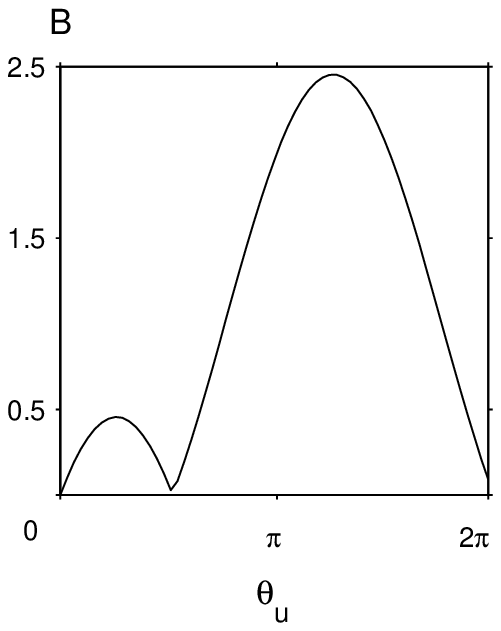} \label{fig2b}}
\end{center}
\caption{Example B}
\end{figure}

In Fig. \ref{fig2a} we show the behavior of the $w_{++}$ 
vs $\theta_1+\theta_2$ for different
values of $n$. Also in this case they 
never exceed the value $1/2$, thus the quantities $E$,
in Eq. (\ref{eq:Edef}), will be always confined between 
$1$ and $-1$, and the Bell's
inequality (\ref{eq:Bdef}) will be satisfied.

For the state (\ref{eq:PsiN}), Eq. (\ref{eq:calExz}) becomes
\begin{equation}\label{eq:EB}
    \mathcal{E}({\bf u}, {\bf v})=
    \left\{
    \begin{array}{ll}
        \cos(\theta_u-\theta_v)  & n = 1, \\
        \cos\theta_u\cos\theta_v & n > 1.
    \end{array}
    \right.
\end{equation}
Then, in Fig. \ref{fig2b} we show $\mathcal{B}$ for the case of $n=1$. 
The nonclassical
character of the entangled state (\ref{eq:PsiN}) becomes now manifest.

\subsection{Example C}

As a third, and last, example we consider a 
pair-coherent state \cite{AGA}, namely
\begin{equation}\label{eq:PC}
    |\psi\rangle = N \int\limits^{2\pi}_0
    |re^{i\varphi}\rangle|re^{-i\varphi}\rangle\,d\varphi,
\end{equation}
where norm $N$ is
\begin{equation}
    N = \frac{e^{r^2}}{2\pi\sqrt{I_0(2r^2)}}
\end{equation}
with $I_{n}$ the modified Besssel function of order $n$. 
The state (\ref{eq:PC}) represents
two coherent states, of the same amplitude $r\in {\bf R}$, 
having a well defined phase
relation although their own phase is random.

The Wigner function corresponding to the state (\ref{eq:PC}) is
\begin{eqnarray}
    &&W(q_1, p_1, q_2, p_2) = 
    \frac{N^2}{\pi^2}e^{-2r^2}\exp(-q^2_1-p^2_1-q^2_2-p^2_2) 
    \times \nonumber \\
    &\times&\int\limits^{2\pi}_0 \int\limits^{2\pi}_0
    \exp\Bigl(\sqrt{2}r(q_1-ip_1)(e^{i\varphi}+e^{-i\varphi'})+
    \sqrt{2}r(q_2-ip_2)(e^{-i\varphi}+e^{i\varphi'})\Bigr)\times 
    \nonumber \\
    &\times&\exp\Bigl(r^2[2\cos(\varphi-\varphi')-\cos(2\varphi)
    -\cos(2\varphi')]\Bigr)\,d\varphi\,d\varphi'.
\end{eqnarray}
The tomogram of the state (\ref{eq:PC}) is
\begin{equation}\label{eq:wPC}
    w(X_1,\theta_1,X_2,\theta_2) = \frac{N^2}{\pi}e^{-2r^2}
    {|I(X_1,\theta_1, X_2,\theta_2)|}^2e^{-X^2_1-X^2_2},
\end{equation}
where the integral $I(X_1,\theta_1,X_2,\theta_2)$ reads
\begin{eqnarray}\label{eq:calI}
    I(X_1,\theta_1,X_2,\theta_2) &=& \int\limits^{2\pi}_0
    \exp\Big[-\frac{r^2}{2}\left(e^{2i(\varphi-\theta_1)}
    +e^{-2i(\varphi+\theta_2)}\right)+\Big. \nonumber \\
    &+&\Bigl.\sqrt{2}r\left(X_1e^{i(\varphi-\theta_1)}+
    X_2e^{-i(\varphi+\theta_2)}\right)\Big]\,d\varphi\,.
\end{eqnarray}
Using the procedure shown in the Appendix together with some algebra, 
we can obtain from Eq.
(\ref{eq:wPC}) the following expressions for the probabilities
$w_{\pm\pm}(\theta_1,\theta_2)$:
\begin{eqnarray}\label{eq:wpmPC}
    w_{\pm\pm}(\theta_1, \theta_2) &=& \frac{N^2}{4}e^{-2r^2}
    \int\limits^{2\pi}_{0}\int\limits^{2\pi}_{0}
    \exp\Bigl(2r^2\cos(\varphi_1+\varphi_2)\Bigr) \times \nonumber \\
    &\times&\left(1\mp{\rm erf}\left(\frac{r}{\sqrt{2}}
    \left[e^{i(\varphi_1-\varphi_0)}+
    e^{i(\varphi_2+\varphi_0)}\right]\right)\right) \times \\
    &\times&\left(1\mp{\rm erf}\left(\frac{r}{\sqrt{2}}
    \left[e^{-i(\varphi_1+\varphi_0)}+
    e^{-i(\varphi_2-\varphi_0)}\right]\right)\right)
    \,d\varphi_1\,d\varphi_2\,, \nonumber
\end{eqnarray}
where ${\rm erf}$ denotes the error function and 
$\varphi_{0}=(\theta_{1}+\theta_{2})/2$.
Notice that the first (second) $\pm$ on the l.h.s. 
corresponds to the first (second) $\mp$
under the integral on the r.h.s.

\begin{figure}
\begin{center}
    \subfigure[The quantity $B$ derived from 
    Eq. (\ref{eq:wpmPC}) is shown vs $r$ for parameters $\theta_u = \pi/2$,
               $\theta_v = -\pi/4$, $\theta_{u'} = 0$, $\theta_{v'} 
               -3\pi/4$.]{\includegraphics{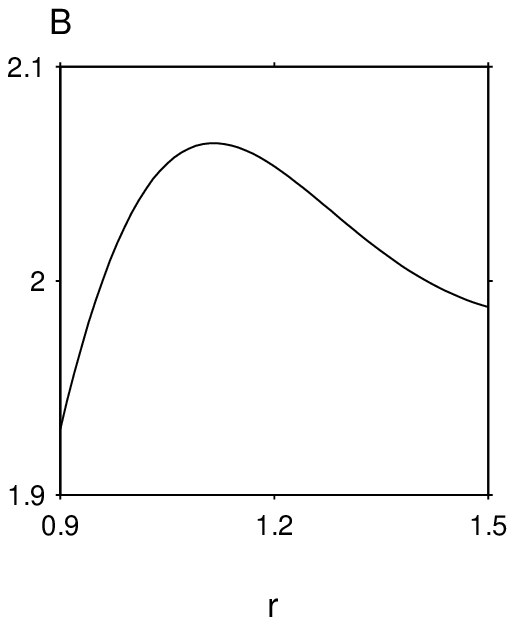} \label{fig3a}}
    \hspace{5mm}
    \subfigure[The quantity ${\cal B}$ derived from Eq.
               (\ref{eq:calEPC}) is shown vs $\theta_u$ 
	       for parameters $r = 1.05$,
               $\theta_{v}=0$, $\theta_{u'}=\pi$, $\theta_{v'}=\pi/2$.]
               {\includegraphics{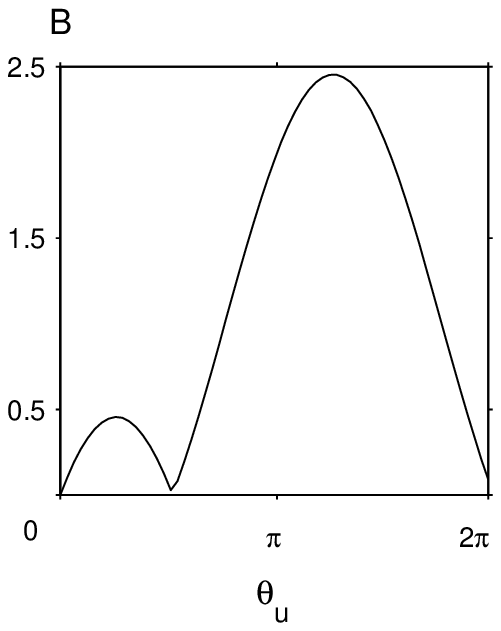} \label{fig3b}}
\end{center}
\caption{Example C}
\end{figure}

The probabilities (\ref{eq:wpmPC}) are no 
longer bounded between the values $\{0,1/2\}$, thus,
they can lead to violation of Bell's inequalities. 
In particular, we show in Fig. \ref{fig3a}
the dependence of $B$ from $r$ revealing 
violation of the CHSH inequality in a small range of
the amplitude $r$. This result recall that of 
Ref. \cite{GIL} where the state (\ref{eq:PC})
was shown to violate another Bell's inequality.

For the state (\ref{eq:PC}), Eq. (\ref{eq:calExz}) becomes
\begin{equation}\label{eq:calEPC}
    \mathcal{E}({\bf u}, {\bf v}) = \cos\theta_u\cos\theta_v +
    r^2\left(1-\frac{J_0(2r^2)}{I_0(2r^2)}\right)
    \sin\theta_u\sin\theta_v\,,
\end{equation}
where $J_{n}$ denotes the Bessel function of order $n$. 
Then, in Fig. \ref{fig3b} we show
$\cal{B}$ for $r=1.05$. The violation of CHSH 
inequality becomes now more pronounced.

\section{Conclusions}

In conclusions we have presented a tomographic approach 
to the quantum nonlocality of a
bipartite continuous variable system. At first instance we have 
proposed a rough use of
tomograms for testing violation of 
local hidden variable theory. In such a case one does not
really need of a complete set of tomograms. However, the 
presence of purely quantum
correlations are not easily recognizable. 
As matter of fact, the EPR state does not show the
nonlocal character, while a pair-coherent state does.

Then, we have proposed to exploit the tomographically reconstructed 
state to get the
statistics of pseudo-spin operators. 
This, practically, maps a continuous variable system into
a discrete variable system making possible the use of 
standard Bell's inequalities. This
seeems a more powerful approach, but 
the price one ought to pay is the large amount of data (a
complete set of tomograms) needed to reconstruct the state. 
In such a case the EPR state
completely shows its nonlocal character. 
Also the pair-coherent state evidenciates much larger
violations \cite{pair}.

Our result opens possibilities for testing QM against 
local hidden variable theories using
very efficient detection methods like homodyne detection \cite{QObook}.

Finally, the present approach could be straightforward 
extended to multipartite systems, e.g.
making possible test of Grenberger-Horne-Zeilinger 
theorem \cite{GHZ} in continuous variables
version \cite{CHEN}. Moreover, it could be useful to 
characterize quantum states used in
information processing \cite{QI}.

\section*{Acknowledgments}
VIM and EVS thank the Russian Foundation for Basic Research for partial
support under projects nos~00-02-16516 and 01-02-17745.

\section*{Appendix}

By refering to Eq. (\ref{eq:calI}) we note that the 
exponent of the imaginary argument is
periodic, and since the integral is taken 
over a period of the exponent we can shift the angle
argument as $\varphi-(\theta_1-\theta_2)/2\to\varphi$ 
so the integral takes the form
\begin{eqnarray}\label{eq:Iint}
    I(X_1,\theta_1,X_2,\theta_2) &=& \int\limits^{2\pi}_0
    \exp\Big[-\frac{r^2}{2}e^{-2i\varphi_0}\left(e^{2i\varphi}
    +e^{-2i\varphi}\right) + \Big. \nonumber \\
    &+&\Big.\sqrt{2}re^{-i\varphi_0}\left(X_1e^{i\varphi}
    +X_2e^{-i\varphi}\right)\Big]\,d\varphi\,,
\end{eqnarray}
where we have set
\begin{equation}
    \varphi_0 = \frac{\theta_1+\theta_2}{2}\,.
\end{equation}
Then, the integral can be rewritten as
\begin{eqnarray}
    I(X_1,\theta_1,X_2,\theta_2) &=& \int\limits^{2\pi}_0
    \exp\left[-\frac{{(\alpha e^{i\varphi})}^2}{2}
    +\sqrt{2}\alpha e^{i\varphi}X_1\right] \times \nonumber \\
    &\times&\exp\left[-\frac{{(\alpha e^{-i\varphi})}^2}{2}+\sqrt{2}\alpha
    e^{-i\varphi}X_2\right]\,d\varphi\,,
\end{eqnarray}
where we have set
\begin{equation}
    \alpha = re^{-i\varphi_0}\,.
\end{equation}
Expanding each exponent in the above integral into a series 
of Hermite polynomials, and taking
into account the equality
\begin{equation}\label{eq:exp}
    \int\limits^{2\pi}_0 e^{i(n-m)\varphi}\,d\varphi = 2\pi\delta_{n,m}\,,
\end{equation}
we obtain the following expression
\begin{equation}\label{eq:I}
    I(X_1,\theta_1,X_2,\theta_2) = 2\pi\sum^{\infty}_{n=0}
    \frac{H_n(X_1)H_n(X_2)}{2^n{(n!)}^2}\alpha^{2n}\,.
\end{equation}

Let us now consider two generic series
\begin{equation}
    f(x_{1}) = \sum\limits^{\infty}_{n=0}f_nx_{1}^n
    \quad {\rm and} \quad
    g(x_{2}) = \sum\limits^{\infty}_{m=0}g_mx_{2}^m\,,
\end{equation}
whose product yields
\begin{equation}
    h(x_{1}, x_{2}) = f(x_{1})g(x_{2})
    =\sum^{\infty}_{n,m=0}f_ng_mx_{1}^nx_{2}^m\,.
\end{equation}
How to express the sum of diagonal elements
\begin{equation}
    s(x_{1}, x_{2}) = \sum^{\infty}_{n=0}f_ng_nx_{1}^nx_{2}^n
\end{equation}
in terms of the $h(x_{1}, x_{2})$ ? 
The answer is found by taking into account the equality
(\ref{eq:exp}), and it reads
\begin{equation}\label{eq:diag}
    s(x_{1}, x_{2}) = \frac{1}{2\pi}\int\limits^{2\pi}_0
    h\left(x_{1}e^{i\varphi}, x_{2}e^{-i\varphi}\right)\,d\varphi\,.
\end{equation}
The sum in right-hand side of Eq. (\ref{eq:I}) 
is the sum of diagonal terms of the product  of
two following series
\begin{equation}
    f_k(\alpha) = \sum^{+\infty}_{n=0}
    \frac{H_n(X_k)}{n!}{\left(\frac{\alpha}{\sqrt{2}}\right)}^n \quad k=1,2.
\end{equation}
If we apply the formula (\ref{eq:diag}) 
to this sum we obtain the representation
(\ref{eq:Iint}) for it which we have started 
from. To obtain probabilities $w_{++}$ etc. we
represent ${|I(X_1, \theta_1, X_2, \theta_2)|}^2$ in the following way:
\begin{equation}
    {|I|}^2 = \int\limits^{2\pi}_0 \int\limits^{2\pi}_0
    f_1(\alpha e^{i\varphi_1})f^*_1(\alpha e^{i\varphi_1})
    f_2(\alpha e^{i\varphi_2})f^*_2(\alpha e^{i\varphi_2})\,
    d\varphi_1\,d\varphi_2.
\end{equation}
To calculate probability $w_{++}$ we need to take the following integral:
\begin{equation}
    \int\limits^{2\pi}_0 \int\limits^{2\pi}_0
    {|I(X_1, \theta_1, X_2, \theta_2)|}^2e^{-X^2_1-X^2_2}\,
    d\varphi_1\,d\varphi_2,
\end{equation}
and analogously for other probabilities. Swap integration 
order in these integrals we obtain
the expression (\ref{eq:wpmPC}).

\end{document}